\documentclass[a4paper,11pt]{article}
\pdfoutput=1 

\usepackage{jcappub} 

\usepackage[T1]{fontenc} 
\usepackage{xspace} 
\usepackage{multirow, makecell}
\usepackage{graphicx}
\graphicspath{{plots/}}

\usepackage[normalem]{ulem}

\newcommand{\nua}[1]{\ensuremath{\rlap{\kern-2.5pt\ensuremath{\overset{\scriptscriptstyle(-)}{\phantom{\nu}}}}{\ensuremath{{\nu}_{#1}}}}\xspace}

\newcommand{\cnb}{C$\nu$B}
\newcommand{\nob}{$N$-one-body}
\newcommand{\Neff}{\ensuremath{N_{\rm eff}}}
\newcommand{\DNeff}{\ensuremath{\Delta N_{\rm eff}}}

\title{\boldmath Calculation of the local density of relic neutrinos}

\author[a]{P.F.\ de Salas,}
\author[a]{S.\ Gariazzo,}
\author[b]{J.\ Lesgourgues,}
\author[a]{and S.\ Pastor}

\affiliation[a]{Instituto de F\'{\i}sica Corpuscular
(CSIC-Universitat de Val\`{e}ncia)\\ 
Parc Cient\'{\i}fic UV, C/ Catedr\'atico Jos\'e Beltr\'an, 2\\ E-46980 Paterna (Valencia), Spain}

\affiliation[b]{Institute for Theoretical Particle Physics and Cosmology (TTK)\\
RWTH Aachen University, D-52056 Aachen, Germany}

\emailAdd{pabferde@ific.uv.es}
\emailAdd{gariazzo@ific.uv.es}
\emailAdd{Julien.Lesgourgues@physik.rwth-aachen.de}
\emailAdd{pastor@ific.uv.es}

\abstract{
Nonzero neutrino masses are required by the existence of flavour oscillations, with values of the order of at least 50 meV.
We consider the gravitational clustering of relic neutrinos within the Milky Way, and used the $N$-one-body simulation technique to
compute their density enhancement factor 
in the neighbourhood of the Earth with respect to the average cosmic density. Compared to previous similar studies, we pushed the simulation down to smaller neutrino masses, and included an improved treatment of the baryonic and dark matter
distributions in the Milky Way. Our results are important for future experiments aiming at detecting the cosmic neutrino background,
such as the Princeton Tritium Observatory for Light, Early-universe, Massive-neutrino Yield (PTOLEMY) proposal.
We calculate the impact of neutrino clustering in the Milky Way on the expected event rate for a PTOLEMY-like experiment.
We find that the effect of clustering remains negligible for the minimal normal hierarchy scenario, while it enhances the event rate by
10 to 20\% (resp.\ a factor 1.7 to 2.5) for the minimal inverted hierarchy scenario (resp.\ a degenerate scenario with 150~meV masses).
Finally we compute the impact on the event rate of a possible fourth sterile neutrino with a mass of 1.3~eV. }
  
\begin{document}
\maketitle
\flushbottom

\section{Introduction}
\label{sec:intro}

The standard cosmological paradigm, known as the Hot Big Bang model, has been very successful in explaining the available cosmological data, but it also 
provides a number of predictions that have not been directly observed yet, such as the presence of a relic sea of neutrinos \cite{NuCosmo}, 
almost as abundant as the photons that constitute the Cosmic Microwave Background (CMB). The existence of this cosmic neutrino background (\cnb) is indirectly 
established by data, in particular the most recent analyses of the power spectrum of CMB anisotropies and other cosmological observables \cite{Ade:2015xua}, 
but its direct detection is hindered by the feebleness of the weak interaction and the smallness of relic neutrino energies, diluted by the gravitational expansion 
since their decoupling time in the early Universe.

The present evidence for flavour neutrino oscillations guarantees that at least two of the three neutrino masses $(m_{1,2,3})$
are not zero, because the squared-mass differences are $\Delta m_{21}^2 \simeq 7.5\times10^{-5}$ eV$^2$ and
$|\Delta m_{32}^2| \simeq 2.5\times10^{-3}$ eV$^2$ \cite{Forero:2014bxa} (see also the more recent global analyses \cite{Capozzi:2017ipn,Esteban:2016qun}),
where $\Delta m_{32}^2>0$ $(<0)$ for normal (inverted) ordering of neutrino masses.
This in turn means that at least two of the neutrino mass eigenstates that form the \cnb~are non-relativistic today, 
since their mass is bigger than their temperature $T_{\nu}^{0} \simeq 1.6\times10^{-4}$~eV.
This represents the only known situation in which neutrinos behave as non-relativistic particles.
Hence, besides being an outstanding achievement for experimental physics
and a further confirmation of the standard cosmological model, the \cnb~detection would allow us to study an unexplored kinematical 
regime \cite{Akhmedov:2017xxm}.

Since the time of the first proposal by Weinberg \cite{Weinberg:1962zza}, several techniques have been studied to detect the relic neutrinos
(see e.g.~\cite{Duda:2001hd,Gelmini:2004hg,Ringwald:2005zf,Li:2015koa,Vogel:2015vfa}), but the task still sounds very challenging.
Given the present tiny values of the neutrino energy, the most promising 
approach is to consider an interaction process with no energy threshold.
In particular, the case of neutrino capture (NC) on $\beta$-decaying nuclei ($\nua{}_e +  A \to e^\pm + A'$) has been considered in refs.~\cite{Cocco:2007za,Lazauskas:2007da,Blennow:2008fh,Faessler:2011qj,Long:2014zva}.
A neutrino capture by a nucleus $A$ that can spontaneously $\beta$-decay stimulates
the emission of an electron with an energy above the $\beta$-decay endpoint.
An experiment based on this process should measure the shape of the energy spectrum of the electrons produced by the $\beta$-decaying
nucleus with exquisite precision near the endpoint. The \cnb~interactions in the detector would be responsible for an energy peak at 
$2m_\nu$ above the $\beta$-decay endpoint. A detection of relic neutrinos could be achieved if the energy resolution $\Delta$ was smaller 
than the neutrino mass. This is a very challenging requirement, because one would need
at least $\Delta\lesssim 0.7m_\nu$ \cite{Long:2014zva} in order to be able to distinguish the events due to neutrino capture from standard $\beta$-decay events.

Among the available beta-decaying nuclei, tritium is considered as the best candidate. This isotope has a 
high neutrino capture cross section, a low Q-value and a long lifetime. Tritium is used as a radioactive source in the KATRIN experiment
\cite{Osipowicz:2001sq,Angrik:2005ep}, whose aim is to determine the absolute neutrino mass by measuring the endpoint region of the
$\beta$ spectrum. KATRIN will start collecting data very soon, but its amount of tritium is far too small for detecting relic neutrinos, with an
estimated event rate of $\mathcal{O}(10^{-6})$ per year \cite{Kaboth:2010kf,Faessler:2016tjf}. On the other hand, 
a dedicated experiment based on neutrino capture by tritium was proposed recently: the 
Princeton Tritium Observatory for Light, Early-universe, Massive-neutrino Yield
(PTOLEMY) \cite{Betts:2013uya}. Its phenomenology and potential for \cnb~detection was studied in detail in ref.\ \cite{Long:2014zva}.
Unfortunately, the designed energy resolution of PTOLEMY, $\Delta\simeq150$ meV, is too large for \cnb\ detection if the heaviest neutrino state has the minimal mass guaranteed by flavour oscillations, of the order of  $m_\nu\simeq 50$ meV. However,
the experiment could be sensitive to larger masses $m_\nu\gtrsim150$ meV, that are 
disfavoured but not completely ruled out by the current cosmological limits on the sum of neutrino masses:
the upper bound at 95\% CL is $\sum m_\nu=m_1+m_2+m_3< 0.34\; (0.17)$ eV 
including Planck CMB temperature and polarization data only (Planck high-$\ell$ temperature + low-$\ell$ polarization + CMB lensing + Baryon Acoustic Oscillations)
\cite{Aghanim:2016yuo}.\footnote{Other cosmological analyses conclude that even the minimal value of $\sum m_\nu$ in the inverted mass hierarchy is disfavoured by some combinations of cosmological data, see e.g.\ ref.\ \cite{Vagnozzi:2017ovm}.
Anyway, one must remember that all these limits are obtained in the context of the $\Lambda$CDM model,
and may change significantly if a different cosmological model is considered.
The changes range from loosen upper limits for simple extensions of the $\Lambda$CDM model (see e.g.\ refs.\ \cite{DiValentino:2016ikp,Capozzi:2017ipn})
to a preference for a positive $\sum m_\nu$ 
when one considers some kind of modified gravity, see e.g.\ refs.\ \cite{Barreira:2014ija,Dirian:2017pwp}.}

For a PTOLEMY-like experiment, working with 100 g of pure atomic tritium, a number of around 10 events per year from \cnb~interactions is expected,
taking into account the present average number density of relic neutrinos \cite{Cocco:2007za,Betts:2013uya,Long:2014zva}. However, massive neutrinos
feel the presence of Dark Matter (DM) halos such as the one of the Milky Way (MW). The potential wells created by DM enhance the neutrino clustering and produce
a higher \cnb\ local density. Therefore, a proper calculation of the overdensity of neutrinos in the Earth's galactic region is important 
to estimate the real number of events that an experiment could observe. Singh \& Ma \cite{Singh:2002de} and Ringwald \& Wong \cite{Ringwald:2004np} have already estimated the overdensity of neutrinos in the MW halo, assuming neutrino masses above $150$ meV. The aim of this paper is to use the same simulation method, the so-called
``\nob{}'' technique described in ref.\ \cite{Ringwald:2004np}, and to improve the calculation of relic neutrino clustering in the local environment.
With respect to \cite{Ringwald:2004np}, we consider lighter neutrino masses, closer to values allowed by the recent cosmological bounds
on the total neutrino mass, in order to obtain more realistic estimates for a PTOLEMY-like experiment. In addition, we also improve the treatment of the matter distribution 
in the local neighbourhood of the Earth, using the results of recent estimates
and $N$-body simulations of MW-like objects for both the DM and baryons densities.

This paper is organised as follows. We describe the method used to compute the gravitational clustering of massive neutrinos in section~\ref{sec:clust}
and how we parameterize the matter distribution (DM and baryons) in our galaxy in section \ref{sec:matter}.
Our results on the local overdensity of massive relic neutrinos and the consequences for PTOLEMY-like experiments are discussed in section \ref{sec:res}.
Finally, we report our conclusions in section \ref{sec:conc}.

\section{Gravitational clustering of massive neutrinos}
\label{sec:clust}

In our work we adopt the ``\nob{}'' simulation method described in ref.~\cite{Ringwald:2004np}
to calculate the clustering of light neutrinos in the local environment.
The \nob{} technique is based on the assumption that the growth of the neutrino overdensity does not influence in a significant way
the evolution of the DM halo and the baryon accretion.
Therefore, it is possible to calculate independently the clustering of each single neutrino in the evolving DM+baryon distribution
and to obtain the total neutrino overdensity as the sum of the contributions of $N$ selected neutrinos.
The crucial difference with respect to an $N$-body simulation is that, instead of evolving $N$ particles at the same time, one evolves $N$ times one single particle.
Thus one can increase arbitrarily the number of sample neutrinos without modifying the complexity of the code.

The \nob{} approach is valid as long as we can assume that:
\begin{enumerate}
\item the only interaction that matters is gravitational;
\item DM and baryons evolve independently of neutrinos: this follows from the small contribution of neutrinos to the total non-relativistic matter density;
\item neutrinos evolve according to the gravitational effects of DM and baryons, and independently of other neutrinos: this is another consequence of the same fact.
\end{enumerate}

\subsection{Equations of motion}\label{ssec:equations}
We can write the Lagrangian for our test neutrino with mass $m_\nu$, moving in a gravitational potential well $\phi(\mathbf{x},\tau)$, as
\begin{equation}
L(\mathbf{x},\mathbf{\dot{x}},\tau) = a \left( \frac{1}{2}m_\nu v^2 - m_\nu \phi(|\mathbf{x}|,\tau) \right),
\end{equation}
where $a=1/(1+z)$ is the cosmological scale factor (normalised to $1$ today), $\mathbf{v}=\mathbf{\dot{x}}$ the peculiar velocity of the particle, 
$\mathbf{x}$ the comoving distance and $\tau$ the conformal time. For simplicity we will choose a spherically symmetric gravitational potential $\phi(\textbf{x},\tau)$, 
so we can rewrite the Lagrangian in comoving polar coordinates $\{ r,\theta \}$,
\begin{equation}
L\left( r,\theta , \dot{r}, \dot{\theta} , \tau \right) = \frac{a}{2} m_\nu \left( \dot{r}^2 + r^2 \dot{\theta}^2 - 2 \, \phi ( r,\tau) \right)\,,
\end{equation}
from where we get the Hamiltonian
\begin{equation}
H\left( r, \theta , p_r , l , \tau \right) = \frac{1}{2 a m_\nu} \left( p_r^2 + \frac{l^2}{r^2} \right) + a m_\nu \phi(r,\tau)\,,
\end{equation}
where
\begin{equation}
p_r = \frac{\partial L}{\partial \dot{r}} = a m_\nu \dot{r},  \qquad \qquad l = r p_\theta = \frac{\partial L}{\partial \dot{\theta}} = a m_\nu r^2 \dot{\theta}
\end{equation}
are the canonical momenta conjugate to $r$ and $\theta$ respectively. We obtain then the Hamilton equations
\begin{align}\label{eq:HamiltonEqs}
&\frac{\partial H}{\partial p_r} = \frac{d r}{d\tau} = \frac{p_r}{a m_\nu}, &&\phantom{-}\;\frac{\partial H}{\partial l} = \frac{d \theta}{	d \tau} = \frac{l}{a m_\nu r^2}, \nonumber \\
-&\frac{\partial H}{\partial r} = \frac{d p_r}{d \tau} = \frac{l^2}{a m_\nu r^3} - a m_\nu \frac{\partial \phi}{\partial r}, &&-\frac{\partial H}{\partial \theta} = \frac{d l}{d \tau} =0,
\end{align}
where the gravitational potential $\phi(r,\tau)$ is known from the Poisson equation
\begin{equation}
\nabla^2 \phi = \frac{1}{r^2}\frac{\partial}{\partial r} \left( r^2 \frac{\partial \phi}{\partial r}\right) = 4\pi G a^2 \rho_{\rm matter}\left(r,\tau\right).
\end{equation}
$G$ is the gravitational constant and we remind the reader that $\rho_{\rm matter}$ is taken to be spherically symmetric. Then we have
\begin{equation}\label{eq:dphidr}
\frac{\partial \phi}{\partial r} = \frac{G}{a r^2} M_{\rm matter}(r,\tau ),
\end{equation}
where
\begin{equation}\label{eq:Mtotal}
M_{\rm matter} (r,\tau) = 4\pi a^3 \int_0^r \rho_{\rm matter} (r' , \tau) r'^2 dr'
\end{equation}
is the total matter at a distance $r$ and a proper time $\tau$.

\subsection{Technical details}

The result of our simulations is a mapping between sets of initial and final neutrino coordinates in phase space.
There are regions in the initial phase space such that today
neutrinos are still inside the dark matter halo, and regions such that they escape and leave to infinity.
Only the first category of trajectories is relevant for us, since our goal is to evaluate the neutrino density in the halo at present time.
However, there is no way to know in advance where the boundaries of the relevant region are in the initial phase space, and we cannot afford to spend most of the computing time on the calculation of irrelevant neutrino trajectories.
We address this problem with an iterative approach. We first launch a set of representative particles for a coarse-grained discretization of the full initial phase space. Among these particles, those ending up inside a certain $r_{\mathrm{cut}}$ radius are traced back, finer discretized and relaunched.
The procedure is repeated until we have a sufficient number of relevant trajectories to estimate the neutrino density profile at $r\simeq r_\odot=8$~kpc with good precision.

The raw result of the simulations consists in a set of final neutrino coordinates in phase space. We need to go from this discrete set to a smooth number density profile $n(r)$. This can be done, first, by assigning an appropriate statistical weight to each trajectory, in order to correctly sample the initial phase space, and to properly take into account the initial Fermi-Dirac isotropic distribution of neutrino momenta; and second, by applying a smoothing kernel to the discrete results, in order to estimate the underlying continuous density distribution.
For these two steps, we follow the procedure described in \cite{Merritt:1994} and already employed in \cite{Ringwald:2004np}.
In particular, since we assume spherical symmetry for simplicity, we must choose a smoothing kernel that will automatically enforce such a symmetry.
Our kernel consists in a Gaussian function of the radial coordinate: in other words, we perform Gaussian smoothing on the surface of spheres of radius $r$. We refer to Appendix A.3 of~\cite{Ringwald:2004np} for the mathematical expressions. 

For each $r$, this kernel has only one arbitrary parameter: the gaussian width $\xi$~(indicated as $h$ in \cite{Ringwald:2004np}).
Choosing a too small $\xi$ would lead to a small number of simulated neutrinos per smoothing shell, and the results would be dominated by shot noise and statistical fluctuations, unless a prohibitive number of trajectories is computed. Choosing a too large $\xi$ would erase the details of the radial density profile $n(r)$ that we want to reconstruct.
Hence the goal of the game is to resolve a sufficient number of trajectories in order to get results which remain independent of $\xi$ within an extended range of $\xi$ values.
As long as the result varies strongly with $\xi$  in the whole range $0<r<r_{\mathrm{cut}}$, we know that the results are not yet converged, and that we need to increase the number of samples.  

In the following results, we use this method to define our convergence criteria. For each simulated model, the optimal value of $\xi$ is chosen in such a way that the results are stable against small variations of $\xi$. The dependence of the result on $\xi$ is used to estimate the numerical error coming from discrete sampling, which is reported as error bands in our final plots. Following this method, we find that smaller neutrino masses require a larger number of neutrino trajectories to achieve the same precision.
This was expected, since lighter neutrinos have larger velocities and escape more easily from the DM halo.
Hence they require a finer discretization of the initial phase space. For the smallest mass considered here, our numerical error is limited by large needs in terms of computing time. However, even in that case, the total errors are dominated by  the uncertainty on the shape of the DM distribution in the MW, rather than by discretization issues and shot noise.

\section{Matter distribution in our galaxy}
\label{sec:matter}

The results of our \nob{} simulations depend on the \emph{total} matter distribution in our galaxy, regardless of its different components.
However, in order to show the relative impact of DM and baryons on the clustering of relic massive neutrinos, we will run simulations with only one component at a time, in addition to full simulations including the whole matter distribution. This will also help us to understand the impact of various approximations and the uncertainty coming from each distribution. 

For simplicity, our simulations do not feature a feedback of baryons on the evolution of DM and vice versa. In the real universe, these feedbacks are important (see e.g.\ \cite{Marinacci:2013mha, Green:2017odb}), but we will see later how we cope with this approximation.
Note also that the DM and baryon profiles implemented here  are estimated from actual observation of the current matter distribution in the MW, which is the result of a self-consistent evolution including feedback effects: hence, we do take feedback effects into account in some indirect way.

\subsection{Dark Matter distribution}
\label{ssec:dm_dist}

For the DM density profile in the MW, we follow  \cite{Pato:2015dua} and use two different assumptions:
\begin{itemize}
\item[a)] A generalised Navarro-Frenk-White (NFW) profile, with a logarithmic slope varying from $\gamma$ below a scale radius $r_s$ to $-3$ in the outer region. We do not fix $\gamma =1$ like in the original NFW proposal, but leave it as a free parameter. Therefore, the NFW profile gets the form
\begin{equation}\label{eq:NFWrho}
\rho_{\rm NFW}(r)= \mathcal{N}_{\rm NFW} \left( \frac{r}{r_s} \right)^{-\gamma} \left( 1 + \frac{r}{r_s} \right)^{-3+\gamma},
\end{equation}
where $\mathcal{N}_{\rm NFW}$ is a normalisation parameter related to the halo mass. Note that there is a unique relation between this parameter and  the DM density at the core radius:
\begin{equation}\label{eq:NFWrho-norm}
\mathcal{N}_{\rm NFW} = 2^{3-\gamma} \rho_{\rm NFW} (r_s).
\end{equation}

\item[b)] An Einasto profile with the parameterisation
\begin{equation}\label{eq:Einrho}
\rho_{\rm Ein}(r) = \mathcal{N}_{\rm Ein} \exp\left\{ -\frac{2}{\alpha}\left( \left( \frac{r}{r_s} \right)^\alpha -1 \right) \right\}.
\end{equation}
The normalisation parameter is now related to the DM density at the core radius by
\begin{equation}\label{eq:Einrho-norm}
\mathcal{N}_{\rm Ein} = \rho_{\rm Ein} (r_s),
\end{equation}
and $\alpha$ is a free parameter in the model.

\end{itemize}

We now fit Milky Way data with our two DM distribution models, described by three parameters
($\mathcal{N}$, $r_s$, $\eta$), where ($\mathcal{N}$, $\eta$) can be either ($\mathcal{N}_{\rm NFW}$, $\gamma$) in the NFW model or ($\mathcal{N}_{\rm Ein}$, $\alpha$) in the Einasto model.

The experimental data are taken from the non-parametric reconstruction of the DM density in the MW performed in \cite{Pato:2015tja}.
In particular, we use the data presented in their figure\ 2, taking the central value of their so-called baryonic bracketing 
as the experimental values for $\rho_{\rm DM}(r)$, and the extreme values as the edge of the $1\sigma$ errors.
Since uncertainties coming from the determination of the galactocentric radius $r$ are included in those on $\rho_{\rm DM}(r)$, we identify $r$ with the central radius of each bin.
Note that \cite{Pato:2015tja} converted angular circular velocities $\omega_{\rm DM}$ to a DM density profile under the assumption of spherical symmetry: this matches perfectly with our own simplifying assumptions. Following the same reference, we assume that the solar radial distance to the galactic centre and the local circular velocity are respectively given by $r_\odot = 8\,\mathrm{kpc}$ and $v_\odot = 230\,\mathrm{km/s}$.

\begin{figure}[t]
\centering
\includegraphics[width=0.7\textwidth, page=1]{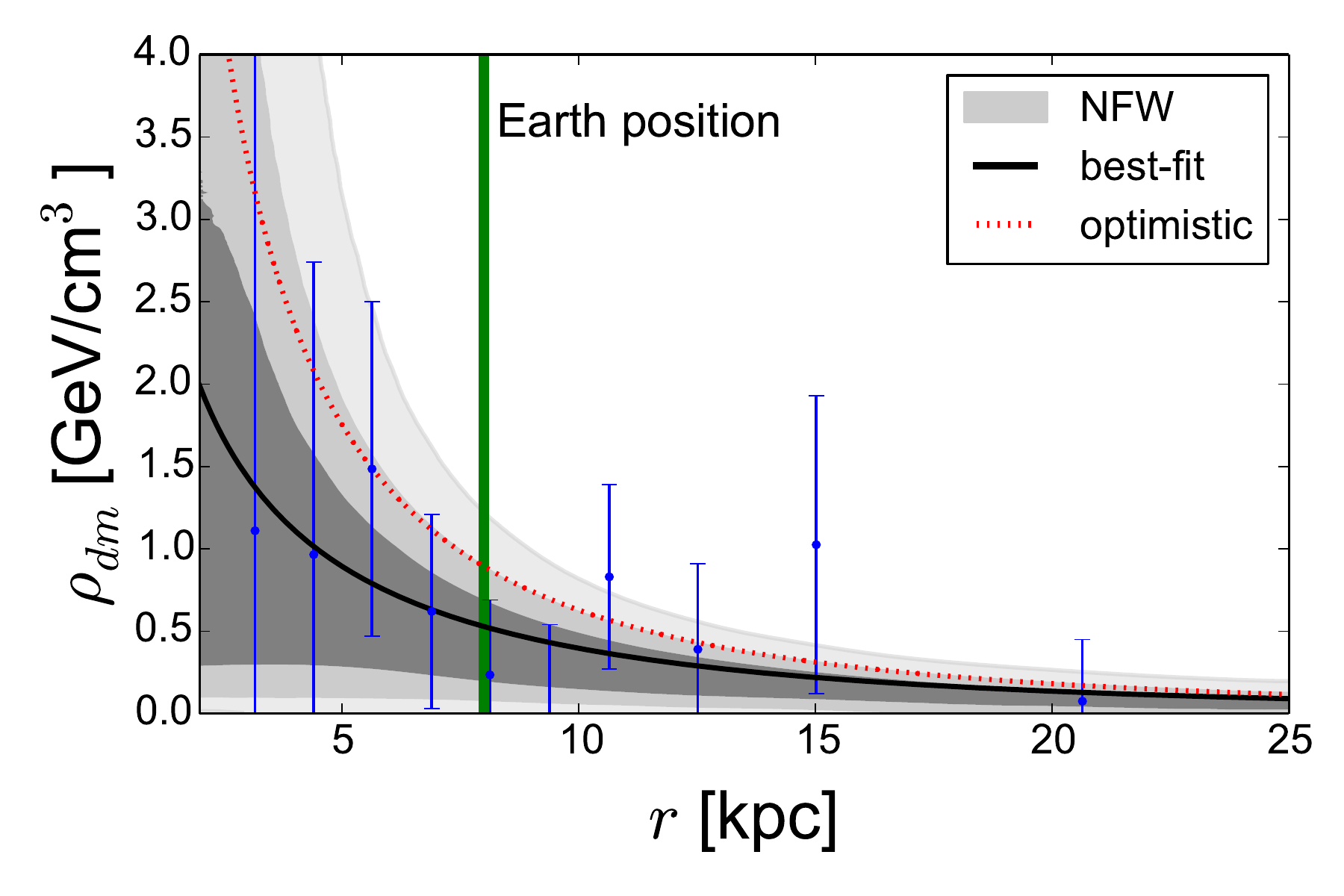}

\includegraphics[width=0.7\textwidth, page=2]{profiles.pdf}
\caption{
Profiles of the dark matter halo that we adopt in the calculations: we show the best-fit and optimistic cases (solid and dotted lines) together with the $1$, $2$ and $3\,\sigma$ regions.
The upper (lower) panel is for a NFW (Einasto) profile of the DM halo.
The green line represents the Earth position.
The blue points are from ref.~\cite{Pato:2015tja}.}
\label{fig:DMprof}
\end{figure}

In order to have an idea of the spread of DM profile parameters in reasonable agreement with the data,
we generated a large amount of possible spectra with a Markov Chain Monte Carlo method.
We used the Metropolis-Hastings to accept or reject models on the basis of their $\chi^2$ statistics.
In each of the two cases (eqs.~\eqref{eq:NFWrho} and \eqref{eq:Einrho}), 
this approach provides the best-fit model (black lines in figure~\ref{fig:DMprof}) and the $1$, $2$ and $3\,\sigma$ confidence intervals (gray bands). We also define in each case what we call the optimistic model (dashed red lines), obtained by picking parameters that nearly saturates the upper $2\,\sigma$ bounds.
We will use these optimistic profiles to obtain an upper limit on the clustering of relic neutrinos at the Sun/Earth radius (shown in green on the figure). 
Table \ref{tab:fit-values} summarises the values of the parameters that describe the best-fit and optimistic NFW and Einasto profiles of figure~\ref{fig:DMprof}, and gives in addition the corresponding virial mass and the parameter $\beta$ (defined and computed below), the density at the Sun/Earth radius, and the $\chi^2$ of the fit to the data points.

\begin{table}
\centering
\resizebox{1\textwidth}{!}{
\begin{tabular}{|l|c|c|c|c|c|c|c|}
\hline
 & $\mathcal{N}$ & $r_s /\mathrm{kpc}$ & $\eta$ & $M_{\rm vir}$/($10^{12}\, \mathrm{M}_{\odot}$) & $\rho_{\rm DM}(r_\odot)$/($\mathrm{GeV/cm^3}$) & $\beta$ & $\chi^2$ \\
\hline
NFW best fit   & $0.73$ & $20.29$ & $0.53$ & $3.76$ & $0.53$ & $2.09$ & $3.06$ \\
\hline
NFW optimistic & $0.73$ & $20.28$ & $0.95$ & $4.25$ & $0.89$ & $2.21$ & $5.79$ \\
\hline
Ein. best fit  & $0.12$ & $20.28$ & $0.45$ & $1.13$ & $0.53$ & $1.10$ & $3.08$ \\
\hline
Ein. optimistic& $0.19$ & $20.27$ & $0.33$ & $2.52$ & $0.94$ & $1.58$ & $6.37$ \\
\hline
\end{tabular}
}
\caption{Values of the parameters for the best-fit and optimistic profiles fitted to data and shown in figure~\ref{fig:DMprof}.}
\label{tab:fit-values}
\end{table}

In order to compute the effect of the gravitational potential $\phi$ on cosmological neutrinos 
through eqs.~\eqref{eq:HamiltonEqs} and \eqref{eq:dphidr} \emph{today}, we need the DM distribution not only at present time, but also in the past.
We assume that the $\eta$ parameter is fixed, while ${\cal N}(z)$ and $r_s(z)$ are functions of the redshift, with values at $z=0$ given by table \ref{tab:fit-values}. We compute the scaling of ${\cal N}(z)$ and $r_s(z)$ with redshift using two constraints (one coming from analytical modelling and one from N-body simulations) on the virial quantities $\Delta_{\rm vir}(z)$ and $c_{\rm vir}(M_{\rm vir},z)$ defined below.

In any spherical halo, the virialized matter forms a sphere with fixed mass $M_{\rm vir}$, varying virial radius $r_{\rm vir}(z)$, and varying overdensity with respect to the critical density $\Delta_{\rm vir}(z)$:
\begin{equation}\label{eq:Mvir-crit}
\Delta_{\rm vir}(z) \equiv \frac{M_{\rm vir}}{\frac{4\pi}{3} a^3 r^3_{\rm vir}(z) \rho_{\rm crit}(z)}~.
\end{equation}
A second obvious relation fulfilled by $M_{\rm vir}$ and $r^3_{\rm vir}$ is
\begin{equation}\label{eq:Mvir-z}
M_{\rm vir} = 4 \pi a^3 \int_0^{r_{\rm vir}(z)} \rho_{\rm DM} (r',z) r'^2 dr'~.
\end{equation}
The function $\Delta_{\rm vir}(z)$ can be inferred from analytic calculations, by following the collapse of a spherical top-hat perturbation~\cite{Bryan:1997dn}, but it is commonly approximated using a fixed value $\Delta\sim200$.
For each of the two parameterisations (NFW and Einasto) we follow the criteria of ref.~\cite{Dutton:2014xda}:
\begin{equation}
\Delta_{\rm vir}(z) =
\begin{cases}
18 \pi^2 + 82 \lambda(z) - 39\lambda(z)^2 \qquad\qquad & \mathrm{for\quad NFW},  \\
200 & \mathrm{for\quad Einasto},
\end{cases} 
\label{eqs:delta}
\end{equation}
where
\begin{equation}
\lambda(z) = \Omega_m(z) - 1
\end{equation}
and $\Omega_m(z)$ is the fractional matter density at redshift $z$.
For each of the four cases studied here (NFW/Einasto with best-fit/optimistic parameters), we know the density profile $\rho_{\rm DM} (r,0)$ today, and $\Delta_{\rm vir}(0)$ is given by  (\ref{eqs:delta}). Then the two equations (\ref{eq:Mvir-crit}, \ref{eq:Mvir-z}) evaluated at $z=0$ provide two relations between two unknowns $M_{\rm vir}$ and $r_{\rm vir}(0)$, that we find numerically.

We now need to relate the virial radius $r_{\rm vir}(z)$ to the scale radius $r_s(z)$ of the NFW or Einasto profile. The ratio
between these two radii is called the concentration parameter:
\begin{equation}\label{eq:cvir-def}
{c_{\rm vir}(M_{\rm vir},z)}={{r_{\rm vir}(z)}/{r_s(z)}}~.
\end{equation}
We use the average concentration parameter across several halos that has been measured in N-body simulations in \cite{Dutton:2014xda}, which provides two functions $a(z)$ and $b(z)$ such that 
\begin{equation}
\log_{10} c_{\rm vir}^{\rm average} = a(z) + b(z)\log_{10}\left( M_{\rm vir}/\left[10^{12} h^{-1} M_\odot \right] \right)~.
\label{eq:cvirav}
\end{equation}
Since this result denotes a statistical trend, we will assume that in each halo $c_{\rm vir}(M_{\rm vir},z) = \beta \times c_{\rm vir}^{\rm average}(M_{\rm vir},z)$, where $\beta$ is a redshift-independent number of order one, that might be different in each galaxy. Knowing $M_{\rm vir}$, $c_{\rm vir}^{\rm average}(M_{\rm vir},0)$, $r_{\rm vir}(0)$ and $r_s(0)$, we can easily compute $\beta$ in the Milky Way for each of our four models.
The $\beta$ values reported in table~\ref{tab:fit-values} are kept fixed in the rest of the calculation.
At this point, the only remaining unknowns are $r_{\rm vir}(z)$, $r_s(z)$ and ${\cal N}(z)$ for $z>0$, but given that $\Delta_{\rm vir}(z)$ and $c_{\rm vir}(M_{\rm vir},z)$ are known from equations (\ref{eqs:delta}, \ref{eq:cvirav}), the three relations (\ref{eq:Mvir-crit}), (\ref{eq:Mvir-z}) and (\ref{eq:cvir-def}) are sufficient for finding them with a numerical algorithm.
Note that this algorithm also depends on the cosmological model through the scaling of the critical density with redshift, involved in equation~(\ref{eq:Mvir-crit}).
During the matter or $\Lambda$ domination, the scaling is given by 
\begin{equation}\label{eq:rhocrit}
\rho_{\rm crit}(z)= \frac{3 H_0^2}{8\pi G}
\left( \Omega_{m,0} (1+z)^3 + \Omega_{\Lambda ,0} \right),
\end{equation}
where $(H_0 , \Omega_{m,0} , \Omega_{\Lambda, 0})$ are the present values of the Hubble parameter, the matter density fraction and the cosmological constant density fraction.
We assume that these parameters take the Planck best-fit values
$(H_0 , \Omega_{m,0} , \Omega_{\Lambda, 0}) = ({67.27~\mathrm{km}/\mathrm{s}/\mathrm{Mpc}, 0.3156, 0.6844})$ \cite{Ade:2015xua}.

\subsection{Baryonic distribution}
\label{ssec:bar}

The baryon distribution in our galaxy is very uncertain. For the DM distribution, we followed ref.~\cite{Pato:2015dua} and discussed only two plausible models. For the baryon distribution, the same reference proposes a total of 70 different models, each of them coming from a different choice of 7 models for the bulge, 5 for the disc and 2 for the gas of the Milky Way.\footnote{See table I in ref.~\cite{Pato:2015dua} for the corresponding list and references.}
Performing \nob{} simulations for each of these cases would require a prohibitively large computation time, and would not bring much to the final results, since the gravitational potential mainly depends on DM.
Actually, in our analysis, the relative uncertainties on the DM distribution correspond to a greater error than the relative uncertainties in the baryon component.

We choose to fix the baryon distribution according to the observation-driven model of ref.~\cite{Misiriotis:2006qq}.
This work fitted simplified axisymmetric distributions to the data for six distinct components: cold and warm dust, molecular and atomic hydrogen, stellar disc and stellar bulge. The latter two distributions are inferred from the star emissivity, which is converted into matter density assuming a global conversion factor determined from a total stellar mass of $6.43\times 10^{10}\,\mathrm{M}_\odot$ \cite{McMillan:2011wd} in the Milky Way.

\begin{figure}[t]
\begin{center}
\includegraphics[width=.9\textwidth]{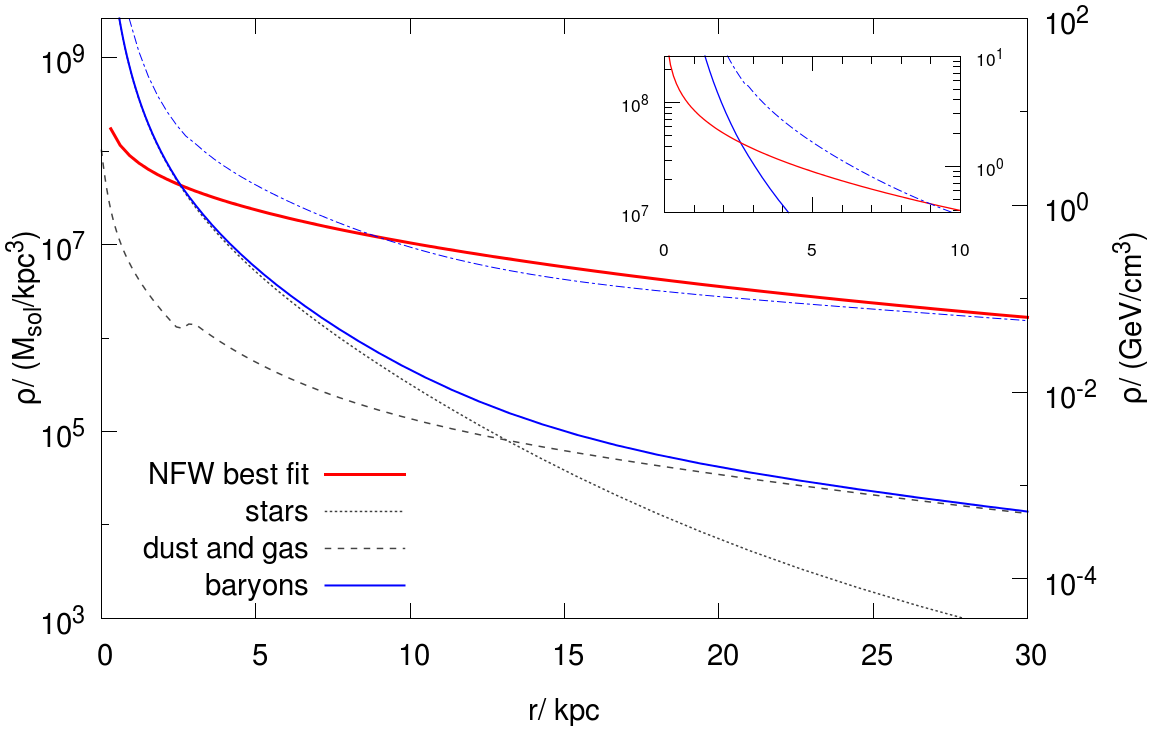}
\caption{Spherically symmetrized baryonic density distribution used in our simulations (blue solid line), and its decomposition in stellar disc contribution
 (dotted) and other components (dashed). We also show the original axisymmetric baryonic profile evaluated on the galactic plane (dotted-dashed, not used in the calculations), and the best-fit NFW DM profile (red line).}
\label{fig:bary-profiles}
\end{center}
\end{figure}

We will further assume that the baryonic profile is spherically symmetric in order to limit the computational time of our simulations. 
This approximation should be harmless for our purposes, since the total effect of baryons on the gravitational potential is smaller than that of the DM, except for a central region of radius $\sim$~5~kpc \cite{Pato:2015tja} dominated by the bulge rather than the disc. 
In order to symmetrize the baryon profile, we simply compute the mass $M_\mathrm{b}(r)$ contained in each sphere  of radius $r$, according to the true axisymmetric distribution. We then take the derivative of this function to obtain the symmetrized baryon density profile.
Our assumptions for the baryonic profile are depicted in figure\ \ref{fig:bary-profiles}, where we show the total baryonic density, both in its spherically symmetrized approximation (solid blue) and on the galactic plane with the original axisymmetric profile (dashed-dotted blue, not used in the calculations). One can see that the two curves are closer at smaller radii, since the thickness of the disc is comparable with the distance from the galactic centre. In order to show that the main contribution to the baryon density comes from the stars, we include different lines for the contribution of stars (dotted black) and the sum of all the other components (dashed black), for the spherically symmetrized profile.
The best-fit NFW DM profile is also shown for comparison.

We model the redshift dependence of the baryon profile through a simple redshift-dependent global normalisation factor ${\cal N}_b(z)$.
For higher precision, one should also introduce a redshift-dependent tilting of the profile, but the impact on our final results would be negligible compared to the uncertainty on the DM profile.
We obtain ${\cal N}_b(z)/{\cal N}_b(0)$ by averaging over the evolution of the stellar mass in eight Milky Way-sized simulated haloes, given in figure\ 18 of ref.\ \cite{Marinacci:2013mha}. 

Concerning the mutual influence that baryons and DM have on each other during the gravitational accretion, it is true that there 
must be a correlation in the profile growth with redshift (see e.g.~\cite{Marinacci:2013mha, Green:2017odb}),
but the effect is small when compared to the uncertainties coming from the shape of the DM profile.
The presence of baryons tends to make the DM halo more clumpy near the galactic centre, but this is already taken into account in our DM profiles since they have been fitted to real data.
At most, we expect that our model slightly overestimates the matter density in the past, because the leading component (DM) is being traced back in time independently, starting from a profile at $z=0$ that includes the current baryonic feedback.

\section{Results and discussion}
\label{sec:res}

In this section we describe our results for the relic neutrino overdensity and the corresponding prospects
for the event rate in a PTOLEMY-like experiment, for different values of the neutrino masses. We 
will first assume nearly minimal neutrino masses
(subsection~\ref{ssec:minimal}), and then consider non-minimal masses, more favourable for a detection by PTOLEMY
(subsection~\ref{ssec:nonminimal}).
Finally, we will show the detection prospects for a light sterile neutrino,
such as the one proposed to solve the short-baseline neutrino oscillation anomalies
(subsection~\ref{ssec:nuster}).

\subsection{Minimal neutrino masses}
\label{ssec:minimal}

\begin{figure}[t]
\centering
\includegraphics[width=0.7\textwidth]{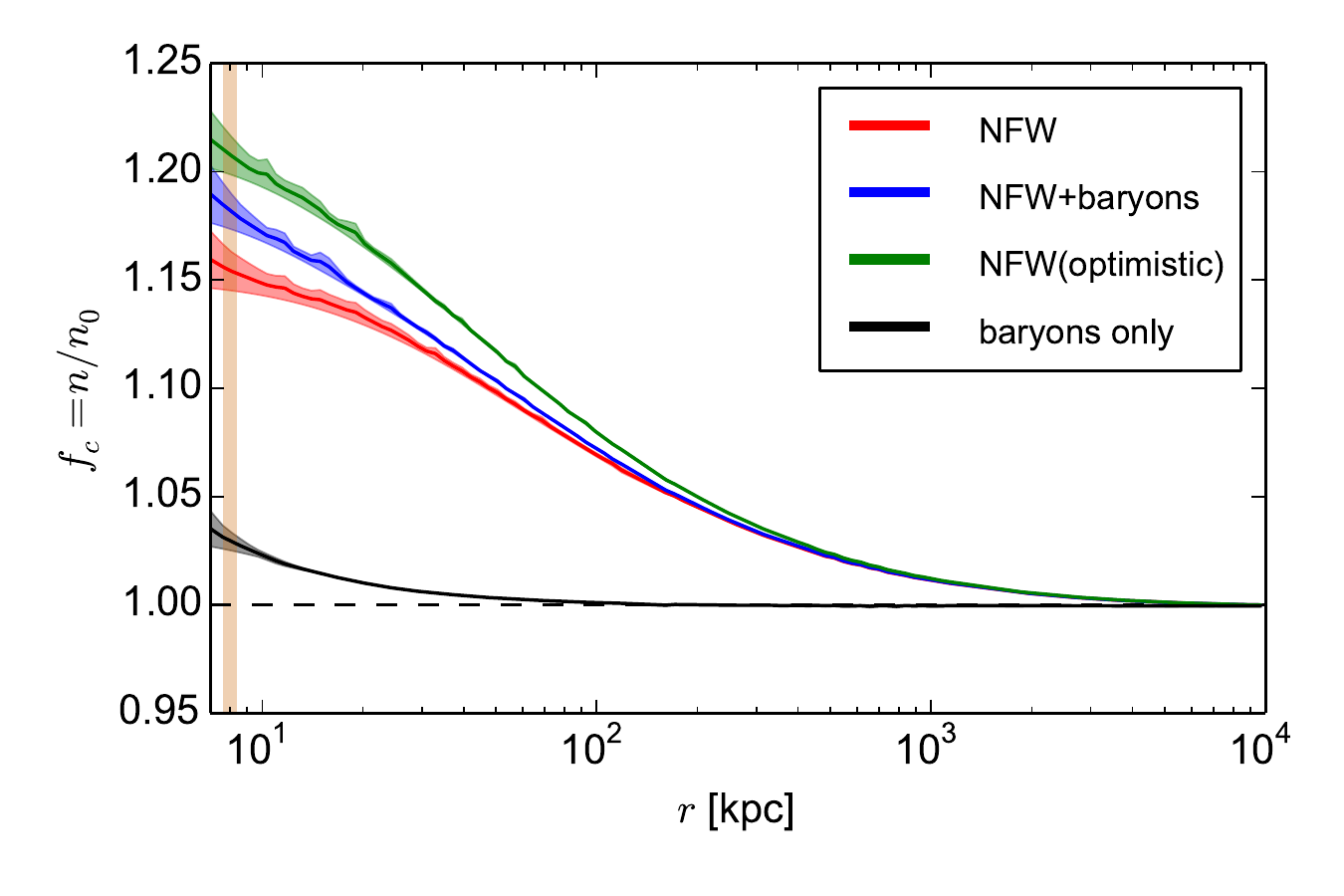}

\includegraphics[width=0.7\textwidth]{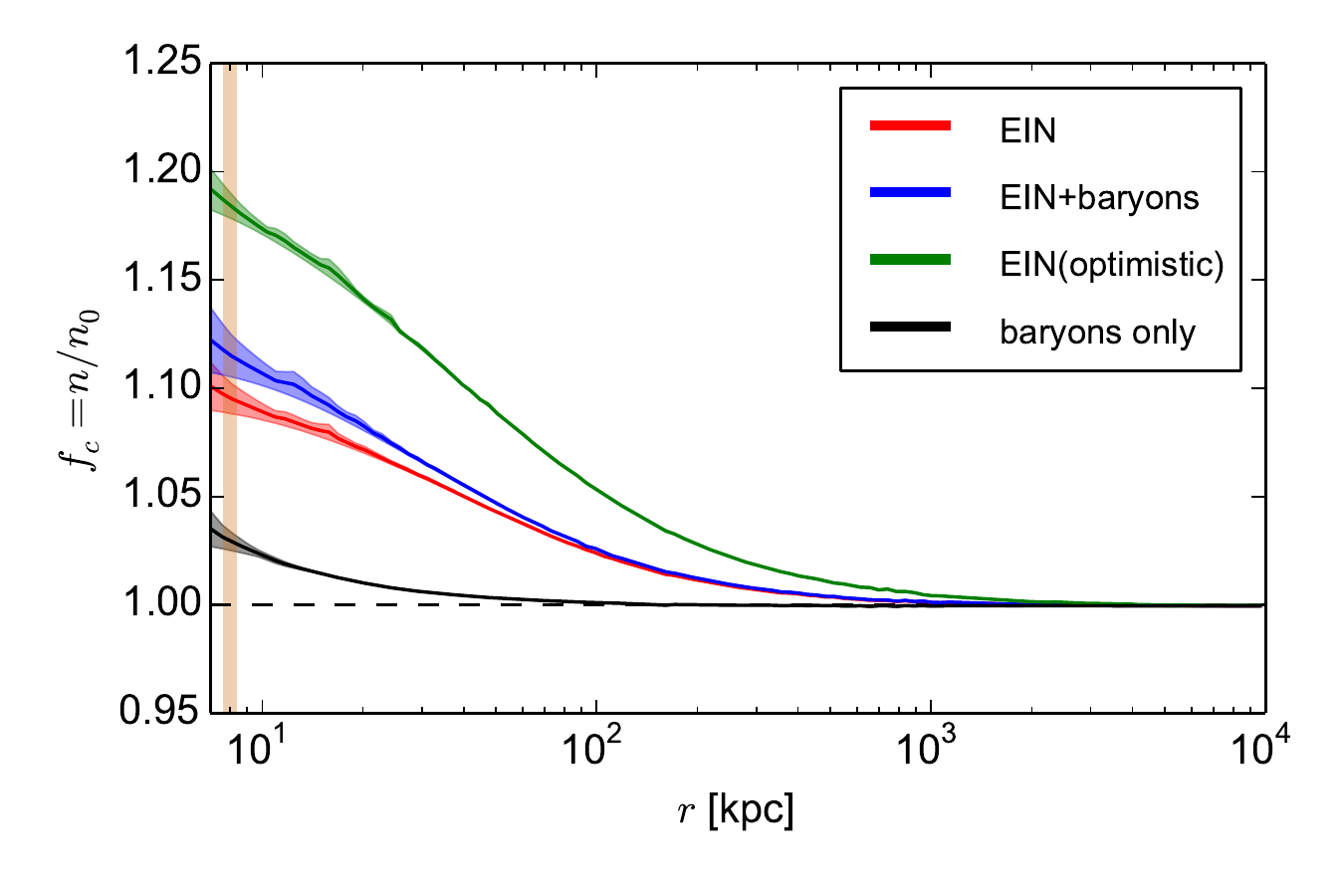}
\caption{
Neutrino overdensity for a single neutrino family with $m_\nu=60$~meV
as a function of the distance from the galactic centre,
computed with different assumptions on the matter profiles.
The upper (lower) panel is for a NFW (Einasto) DM profile.
The orange band represents the Earth position.}
\label{fig:results60meV}
\end{figure}

Using the method described in the previous sections, we show here the results obtained for nearly minimal neutrino masses,
when the heaviest mass eigenstate has a mass $m_\nu\simeq 60$~meV. We show in figure\ \ref{fig:results60meV} the overdensity profile 
of such a neutrino state, for the different assumptions on the DM and baryon distribution discussed previously. For each
DM profile, NFW or Einasto, we consider the best-fit case and the optimistic case (described in Sec.\ \ref{ssec:dm_dist} and listed in table \ref{tab:fit-values}). We also show the effect of the baryonic component of section~\ref{ssec:bar}, alone or combined with each best-fit case.

For a neutrino with $m_\nu\simeq 60$~meV, we find that the 
total overdensity due to gravitational clustering at the Earth distance from the galactic centre is rather small,
with a relative increase with respect to the average density of the \cnb~of 10-20\% at most. 
The values of the local neutrino overdensity are reported in table \ref{tab:rates60} for the different matter profiles.

For the same normalisation of $\rho_\mathrm{DM}(r)$ at the Earth distance, the NFW profile corresponds to a higher DM density than the Einasto profile at the galactic centre, and consequently to a stronger gravitational attraction of neutrinos.
Thus, one can see in figure\ \ref{fig:results60meV} that the local overdensity generated by a NFW-distributed DM is larger than in the Einasto cases. As expected, the effect of baryons is small, modifying the result by approximately 3\% with respect to the \cnb{} mean density.

These numbers slightly depend on the numerical factor $\xi$ used in the reconstruction of the neutrino profile, and the corresponding
numerical uncertainty is represented by the coloured band that enclose each curve in figure\ \ref{fig:results60meV}.
These bands do not represent any kind of statistical or systematic error in the whole calculation:
they just quantify the difficulty
of robustly computing the neutrino clustering at decreasing distances from the centre of the matter halo.

The final error is dominated by the  uncertainty on the DM profile, since 
there is a significant difference between the NFW and the Einasto cases, and between the best-fit and optimistic models in each case. We recall that our \emph{optimistic} models can be considered as upper limits on the DM distributions,
since they are chosen to be overestimates of the actual available data. 

One can also see from figure\ \ref{fig:results60meV} that the overdensity, as expected, depends on the distance from the centre of the MW.
We see that neutrinos feel the gravitational attraction of the Milky Way halo and cluster at radii $r \leq r_{\rm over}\simeq\mathcal{O}(1\text{~Mpc})$.

The neutrino overdensity on sub-galactic scales was not previously computed for values of the neutrino mass as small as $60$ meV. Therefore,
we cannot directly compare our results with other similar works in the literature.
However, when comparing with the \nob{} results of ref.\ \cite{Ringwald:2004np} and the
full N-body results of ref.\ \cite{VillaescusaNavarro:2012ag} obtained for heavier neutrinos, we find that the shape and the normalisation of our profiles are compatible with their results.

We must comment the fact that the calculations we performed for such a (nearly) minimal value of the neutrino mass can be developed using the linear approximation with minor changes in the final results.
The differences between the two methods have been studied in ref.~\cite{Ringwald:2004np}, where it is shown that the approximated results are very close to the simulated ones if the mass of the galaxy and the neutrino mass are small.
However, since we also want to compute the overdensity for heavier neutrinos, it is mandatory to use the full computation.

In our \nob{} simulation, we only considered one isolated spherical galactic halo with homogeneous boundary conditions, 
and neglected the effect of neighbouring galaxies, satellite galaxies, etc. The impact of 
these objects should be negligible at $z=3$, when we start the simulation, but it could become important near $z=0$.
However, the computation of the neutrino clustering in a realistic environment
where all DM clumps close to the MW are properly represented would be extremely demanding, so here
we just present a qualitative discussion of their possible effects. 

The satellite galaxies located at less than a Mpc from the centre of the Milky Way (the distance 
at which the DM halo starts to influence the neutrino overdensity) are much lighter than the
MW, so we expect that their gravitational effect will be very small. On the other hand, the Andromeda galaxy is 
slightly larger than the MW and it is only $\sim0.8$ Mpc away (see e.g.\ \cite{McConnachie:2004dv,Penarrubia:2014oda}).
If we simply consider the superposition of two distinct (independently evolved) neutrino haloes for the two galaxies,
we conclude from our results that the overdensity $f_\mathrm{c}$ at the Earth increases by $\sim 0.02-0.03$ due to Andromeda: this is comparable to the effect of baryons, and smaller than the uncertainties due to the DM halo. For more distant galaxies with size comparable to the MW the effect would be even smaller. However, we need to take in consideration also larger objects, and in particular the Virgo cluster, the closest galaxy cluster to the MW.
Its centre is located at a approximately $16$~Mpc, with a mass around 
$10^{15}M_\odot$ \cite{Mei:2007xs,Fouque:2001qc}.
Interestingly, the neutrino clustering in a DM halo comparable to the one of the Virgo cluster
has been studied with a full N-body simulation in ref.\ \cite{VillaescusaNavarro:2011ys},
for neutrino masses ranging from 50 to 300 meV. One can see in their figure\ 2 
that the Virgo neutrino halo probably extends beyond $10$~Mpc. An extrapolation of the curve suggests 
that for a mass close to 50 or 60~meV, Virgo may create a neutrino overdensity at a radius of $16$~Mpc
of the same order as the one created by the MW halo itself.
Therefore, the clustering factor due to Virgo should be kept into account when computing the local \cnb{} neutrino density, but
a full calculation involving different DM halos in the neighbourhood of the MW would be computationally too expensive
to be performed with our method.

Let us now consider the effect of an enhanced local density of \cnb{} neutrinos on the expected interaction rate in a future detector.
Following \cite{Long:2014zva}, this can be estimated from
\begin{equation}
\label{eq:gamma_tot_osc}
 \Gamma_{\text{C}\nu\text{B}}^{}
 =
 \sum_{i=1}^{3}
 |U_{ei}|^2[n_i(\nu_{h_R})+n_i(\nu_{h_L})]\,N_T\,\bar\sigma
\end{equation}
where $n_i(\nu_{h_{R(L)}})$ is the number density of the $i$th mass eigenstate neutrino with right (left) helicity,
$N_T=M_T/m(^3\mathrm{H})$ is the approximate number of tritium nuclei in a sample of mass $M_T$,
and $\bar\sigma\simeq3.834\times10^{-45}\text{ cm}^2$ \cite{Long:2014zva}.
In the following we will assume that the number density in the vacuum of a single neutrino in a given helicity state is $n_0$, for all neutrinos, and independently of the mass.
We use the values $|U_{ei}|^2=(0.681, 0.297, 0.022)$ \cite{Olive:2016xmw}
for the mixing matrix elements, and we fix the tritium mass to
$M_T=100$~g, having in mind the PTOLEMY proposal \cite{Betts:2013uya}.

If the local density of each mass eigenstate is increased by a clustering factor $f_c(m_i)$,
the total neutrino capture rate can be obtained as
\begin{equation}
\label{eq:gamma_tot_osc_clus}
 \Gamma_{\text{C}\nu\text{B}}^{}
 =
 [n_0(\nu_{h_R})+n_0(\nu_{h_L})]\,N_T\,\bar\sigma
 \sum_{i=1}^{3}
 |U_{ei}|^2f_c(m_i)\,.
\end{equation}
We recall that the capture rate of Majorana neutrinos $\Gamma_{\text{C}\nu\text{B}}^{M}$
is twice the one of Dirac neutrinos $\Gamma_{\text{C}\nu\text{B}}^{D}$,
because the capture is kinematically forbidden for right-helical Dirac antineutrinos,
while in the Majorana case both helicity states interact weakly.
As a reference, we will compare our results with the ones obtained when neglecting the neutrino clustering due to the MW \cite{Long:2014zva},
\begin{equation}
\label{eq:gamma_DM}
 \Gamma_{\text{C}\nu\text{B}}^{D}
 \simeq4.06\text{ yr}^{-1}\,,
 \qquad
 \Gamma_{\text{C}\nu\text{B}}^{M}
 =
 2 \Gamma_{\text{C}\nu\text{B}}^{D}
 \simeq8.12\text{ yr}^{-1}\,.
\end{equation}

In table \ref{tab:rates60} we summarise our results for the normal and inverted hierarchy scenarios,
assuming each time that the heaviest neutrino mass is $\simeq60$~meV.
We can see that the choice of mass ordering matters for computing the capture rate,
since the three mass eigenstates are differently mixed with the electron neutrino flavour, the only one that interacts with the detector nuclei.
In the case of normal mass ordering,
gravitational clustering has practically no effects on the final event rate.
For inverted ordering, there is a noticeable increase of the event rate of the order of 10 to 20\%.

We conclude this discussion recalling that, unfortunately, a higher event rate does not necessarily lead to an easier detection of the \cnb{}.
Depending on the detector resolution, the events due to interactions with the C$\nu$B may be distinguished or not
from those coming from the standard $\beta$-decay of the detector material, that constitutes the main background for these kind of experiments.
Since a resolution $\Delta\lesssim0.7m_\nu$ is required for a neutrino mass $m_\nu$,
while the PTOLEMY experiment targets $\Delta \sim 100-150$~meV,
the cases that we have discussed so far are beyond the capabilities of a PTOLEMY-like experiment.

\begin{table}[t]
\centering
\resizebox{1\textwidth}{!}{
\begin{tabular}{c|c||c|c|c||c|c}
\multirowcell{ 2}{masses} &
\multirowcell{ 2}{ordering} &
\multirowcell{ 2}{matter halo} &
\multicolumn{2}{c||}{overdensity $f_c$} &
\multirowcell{ 2}{$\Gamma_{\rm tot}^D$~(yr$^{-1}$)}  &
\multirowcell{ 2}{$\Gamma_{\rm tot}^M$~(yr$^{-1}$)} \\
& & & $f_1\simeq f_2$&$f_3$ & & \\
\hline\hline
any & any & any & \multicolumn{2}{c||}{no clustering} & 4.06 & 8.12 \\
\hline
\multirowcell{ 4}{
$m_3=60$~meV}
& \multirowcell{ 4}{NO} & NFW(+bar)      & \multirowcell{4}{$\sim$1} & 1.15 (1.18) & 4.07 (4.08) & 8.15 (8.15) \\
&                       & NFW optimistic &                     & 1.21        & 4.08        & 8.16        \\
&                       & EIN(+bar)      &                     & 1.09 (1.12) & 4.07 (4.07) & 8.14 (8.14) \\
&                       & EIN optimistic &                     & 1.18        & 4.08        & 8.15        \\
\hline
\multirowcell{ 4}{$m_1\simeq m_2=60$~meV}
& \multirowcell{ 4}{IO} & NFW(+bar)      & 1.15 (1.18) & \multirowcell{4}{$\sim$1} & 4.66 (4.78) & 9.31 (9.55) \\
&                       & NFW optimistic & 1.21        &                     & 4.89        & 9.77        \\
&                       & EIN(+bar)      & 1.09 (1.12) &                     & 4.42 (4.54) & 8.84 (9.07) \\
&                       & EIN optimistic & 1.18        &                     & 4.78        & 9.55        \\
\hline
\end{tabular}
}
\caption{
Clustering factors and expected event rates for Dirac or Majorana neutrinos in a PTOLEMY-like experiment,
under different assumptions on the matter profile and on the neutrino mass ordering,
when the heaviest mass is 60~meV. The first line shows the event rates in absence of clustering (for an homogeneous C$\nu$B).
}
\label{tab:rates60}
\end{table}

\subsection{Active neutrinos with non-minimal masses}
\label{ssec:nonminimal}

\begin{figure}[t]
\centering
\includegraphics[width=0.7\textwidth]{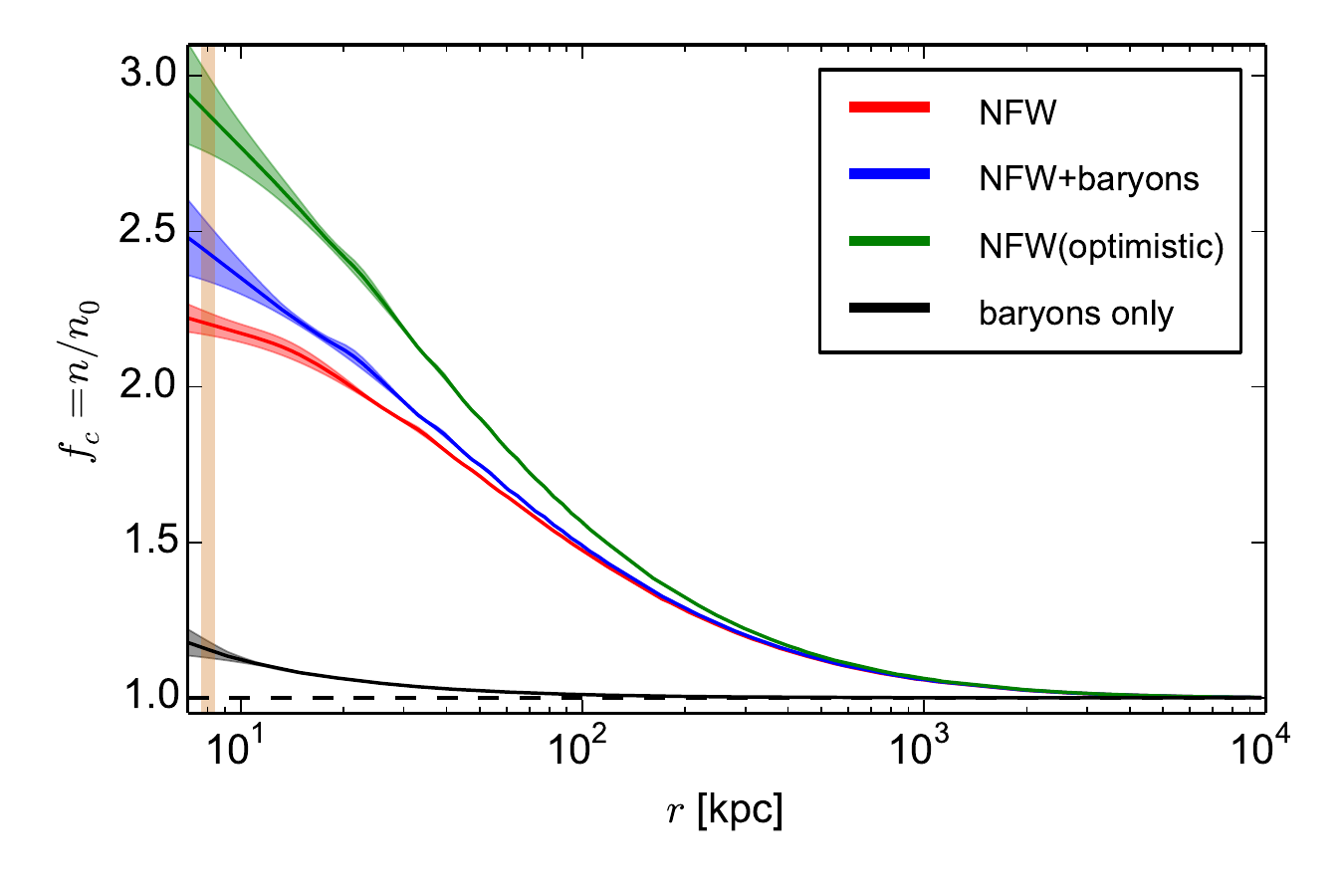}
\includegraphics[width=0.7\textwidth]{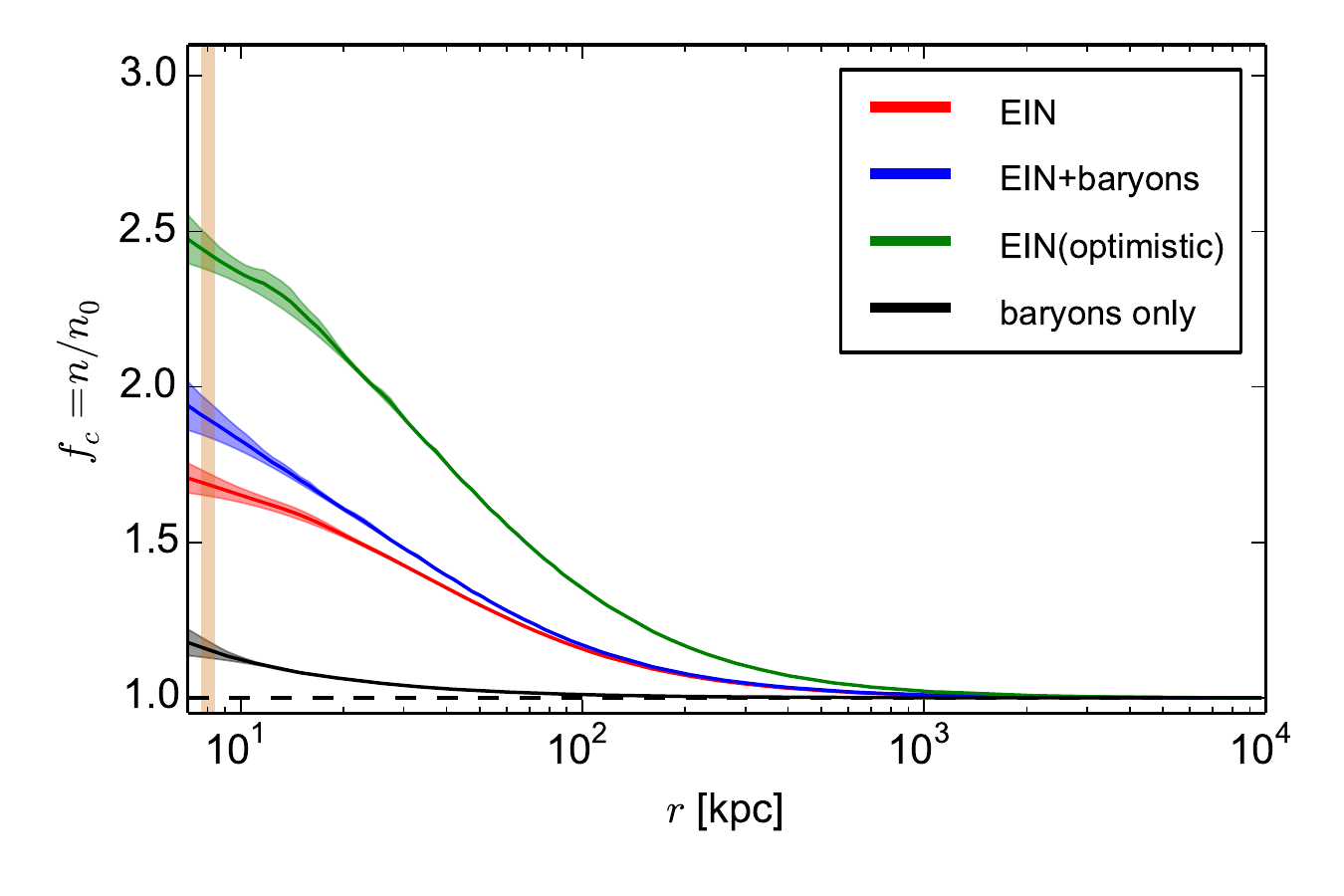}
\caption{
Same results as Figure \ref{fig:results60meV}, but for a neutrino mass $m_\nu=150$ meV.}
\label{fig:results150meV}
\end{figure}

After considering the most pessimistic neutrino mass scenarios from the point of view of C$\nu$B detection, we turn to a very optimistic 
scenario in which the heaviest neutrino would have a mass  $m_\nu\simeq150$ meV. This assumption is motivated by the expected resolution of the PTOLEMY experiment, $\Delta\simeq100-150$~meV \cite{Betts:2013uya}.
In such a case, both the squared-mass differences probed by oscillation experiments are much smaller than $m_\nu^2$, and the three neutrino states share practically the same mass $m_\nu$.
In such a degenerate scenario, in which the mass ordering is not so relevant,
the event rate is 
directly proportional to the overdensity factor $f_c$ and the total neutrino mass is
$\sum m_\nu\simeq 450$ meV, a value already excluded by the most constraining combinations of cosmological data in the minimal $\Lambda$CDM framework (see e.g.\ \cite{Ade:2015xua,Vagnozzi:2017ovm}),
but still plausible in extended models (see e.g.\ refs.\ \cite{DiValentino:2016ikp,Capozzi:2017ipn,Barreira:2014ija,Dirian:2017pwp}).

We repeated our  \nob{} simulations for 150~meV neutrinos. Our results are summarised in figure\ \ref{fig:results150meV}
and table \ref{tab:rates150}, using the same format as in the previous figures and tables.
The general trends are the same as for 60~meV neutrinos, but the local overdensity factor is much larger: it reaches values around 2.4 (1.9),
with a maximum of 2.9 (2.4) in the optimistic NFW (Einasto) case. As in the previous case, we can see that the neutrino halo of the Milky Way extends up to $r\simeq1$ Mpc.

Our results can be compared with those presented in ref.\ \cite{Ringwald:2004np} for the same neutrino mass.
Since we are considering a different value for the DM halo mass, a direct comparison with their figure~5 is not possible,
but our profiles are in good agreement with their results in figure~1.
The same can be said about figure~8 of ref.~\cite{VillaescusaNavarro:2012ag}:
there is a qualitative agreement, but the different DM halo masses prevent a direct comparison.

Concerning the direct detection of relic neutrinos in a PTOLEMY-like experiment, table \ref{tab:rates150} shows
that for a neutrino mass
of order $150$~meV,
gravitational clustering in the Milky Way halo may enhance the expected event rates by a factor 1.7 to 2.5 depending on the DM profile.
Under the most optimistic assumptions and for Majorana neutrinos, one may expect more than 20 events per year.

\begin{table}[t]
\centering
\begin{tabular}{c|c||c|c}
\multirowcell{ 2}{matter halo} &
overdensity $f_c$ &
\multirowcell{ 2}{$\Gamma_{\rm tot}^D$~(yr$^{-1}$)}  &
\multirowcell{ 2}{$\Gamma_{\rm tot}^M$~(yr$^{-1}$)} \\
& $f_1\simeq f_2\simeq f_3$ & & \\
\hline\hline
any & no clustering & 4.06 & 8.12 \\
\hline
NFW(+bar)      & 2.18 (2.44) & 8.8 (9.9) & 17.7 (19.8) \\
NFW optimistic & 2.88        & 11.7      & 23.4        \\
EIN(+bar)      & 1.68 (1.87) & 6.8 (7.6) & 13.6 (15.1) \\
EIN optimistic & 2.43        & 9.9       & 19.7        \\
\hline
\end{tabular}
\caption{
Clustering factors and expected event rates in a PTOLEMY-like experiment
for different assumptions on the matter profile,
when neutrinos are approximately degenerate in mass and $m_\nu\simeq150$~meV.
}
\label{tab:rates150}
\end{table}

\subsection{Beyond active neutrinos: light sterile neutrinos}
\label{ssec:nuster}

A light sterile neutrino is a proposed extension of the Standard Model of particle physics, with the
aim of explaining the so-called short-baseline (SBL) neutrino oscillations anomalies. 
Within a three-flavour neutrino scheme, it seems challenging to explain 
the LSND \cite{Athanassopoulos:1995iw,Aguilar:2001ty} data,
the Gallium \cite{Abdurashitov:2005tb,Laveder:2007zz,Giunti:2006bj,Giunti:2010zu,Giunti:2012tn} anomaly
and the reactor \cite{Mention:2011rk} anomaly.  
Beyond systematic uncertainties, a possible explanation would involve oscillations with
a fourth neutrino mass eigenstate, requiring a new sterile state having a small mixing with the three active neutrino flavours
(see also refs.~\cite{Bilenky:1998dt,GonzalezGarcia:2007ib,Conrad:2012qt,Gariazzo:2015rra}).

In the framework of the so-called 3+1 active-sterile mixing scheme,
the fourth mass eigenstate is heavier than the standard three ones,
with a squared-mass difference
$\Delta m^2_{\rm SBL}
=\Delta m^2_{41} \simeq \Delta m^2_{42} \simeq \Delta m^2_{43}
\gtrsim 1 \, \text{eV}^2
\gg |\Delta m^2_{32}|
\gg \Delta m^2_{21}$.
The $3\times3$ mixing matrix is extended to a $4\times4$ mixing matrix $U$.%
\footnote{We will use the same name for the $3\times3$ and the $4\times4$ mixing matrix,
since $U$ refers to the $4\times4$ only in this section and there is no confusion.}
Its elements in the fourth column quantify the mixing of the fourth mass eigenstate, $\nu_4$,
with the active neutrino flavours. Experimental data impose that they must be small: $|U_{i4}|^2 \ll 1 $, where $i=e,\,\mu,\,\tau$.
Therefore the standard three neutrino mixing is not affected by the existence of the new mass eigenstate.

In such a scheme, the new neutrino oscillates with the active flavours in the early Universe and
a population of $\nu_4$ may be created, depending on the values of the new mixing parameters.
We consider the usual parameterisation of the energy density of radiation
in the early Universe in terms of the effective number of neutrinos \Neff,
\begin{equation}\label{eq:neff}
\rho_r=\left[1+\frac{7}{8}\left(\frac{T_\nu}{T_\gamma}\right) \Neff\right] \rho_\gamma\;,
\end{equation}
where $\rho_r\,(\rho_\gamma)$ is the total energy density of relativistic species (of photons).
The factor $7/8$ accounts for the fermionic degrees of freedom,
while the ratio $T_\nu/T_\gamma=(4/11)^{4/3}$  corresponds to the
difference in the temperatures of cosmological neutrinos and photons in the limit of instantaneous neutrino decoupling.
The effective number is $\Neff^{\rm active}=3.045$~\cite{Mangano:2005cc,deSalas:2016ztq} in presence of active neutrinos only,
and the additional contribution given e.g.\ by the sterile neutrino can be written as
\begin{equation}
\label{eq:dneff}
\DNeff=
\left[\frac{1}{\pi^2}\int dp\; p^3 f_s(p)\right] / \left[\frac{7}{8}\frac{\pi^2}{15}{T_\nu}^4\right] = \Neff-3.045\;,
\end{equation}
where $f_s(p)$ is the energy distribution function of the sterile neutrino in terms of its momentum $p$.

The parameter \DNeff{} is crucial to know how efficiently the fourth neutrino was created in the early Universe and,
as a consequence, gives its contribution to the \cnb{}. Its value depends on $f_s(p)$, which in turn is
fixed by the production mechanism of the sterile neutrinos. The simplest possibility is that the $\nu_4$'s
were generated by active-sterile oscillations in the early Universe
\cite{Dolgov:2003sg,Cirelli:2004cz,Melchiorri:2008gq,Hannestad:2012ky,Mirizzi:2013gnd,Hannestad:2015tea}
with the same temperature (and momentum distribution) of active neutrinos.
In this case, considering the current best-fit results on active-sterile neutrino oscillation parameters,
we would have $\DNeff=1$, that is incompatible with the most recent CMB determinations \cite{Ade:2015xua}.
Instead, if the production of sterile neutrinos occurs through non-thermal mechanisms,
the fourth neutrino momentum distribution maintains approximately the same shape of the 
active neutrino ones \cite{Dodelson:1993je,Acero:2008rh,Jacques:2013xr}
and can be written as a Fermi-Dirac spectrum times a constant scaling factor, that is \DNeff{}.
Hence, in the following we will assume
\begin{equation}\label{eq:fs_p}
 f_s(p) = \frac{\DNeff}{1+\exp(p/T_\nu)}\,
\end{equation}
as an input for the calculation of the local overdensity of the fourth neutrino.
As the mean neutrino number density today is defined by
\begin{equation}\label{eq:numdens}
\bar n_i = \frac{g_i}{(2\pi)^3}\int f_i(p)\, d^3p\,,
\end{equation}
for the fourth neutrino we have $\bar n_4=n_0\,\DNeff$.
This number will be multiplied by the clustering factor $f_c(m_4)$, which is independent of $\DNeff$, in order to obtain
the local number density of $\nu_4$, $n_4 = n_0\,f_c(m_4)\,\DNeff$, relevant for \cnb~detection.

In order to compute the expected number of events in PTOLEMY from the additional neutrino state,
we must include the relevant element of the fourth column of the mixing matrix, $|U_{e4}|^2$.
This parameter and the SBL squared-mass difference $\Delta m^2_{\rm SBL}$ are the only quantities needed to estimate
the event rate in PTOLEMY. Both values should be obtained from a global fit of all neutrino oscillation data.
The most recent global fit that considers the anomalous SBL experiments
reports the best fit values
$\Delta m^2_{\rm SBL}=1.7$ eV$^2$ and 
$|U_{e4}|^2=0.020$ \cite{Gariazzo:2017fdh} for the parameters we are interested in.
In particular, these results were obtained taking into account the
recent measurements of the NEOS \cite{Ko:2016owz} experiment,
that provide a $\sim2\sigma$ hint of the existence of SBL oscillations while slightly lowering
the best-fit value of $|U_{e4}|^2$.

\begin{figure}[t]
\centering
\includegraphics[width=0.7\textwidth]{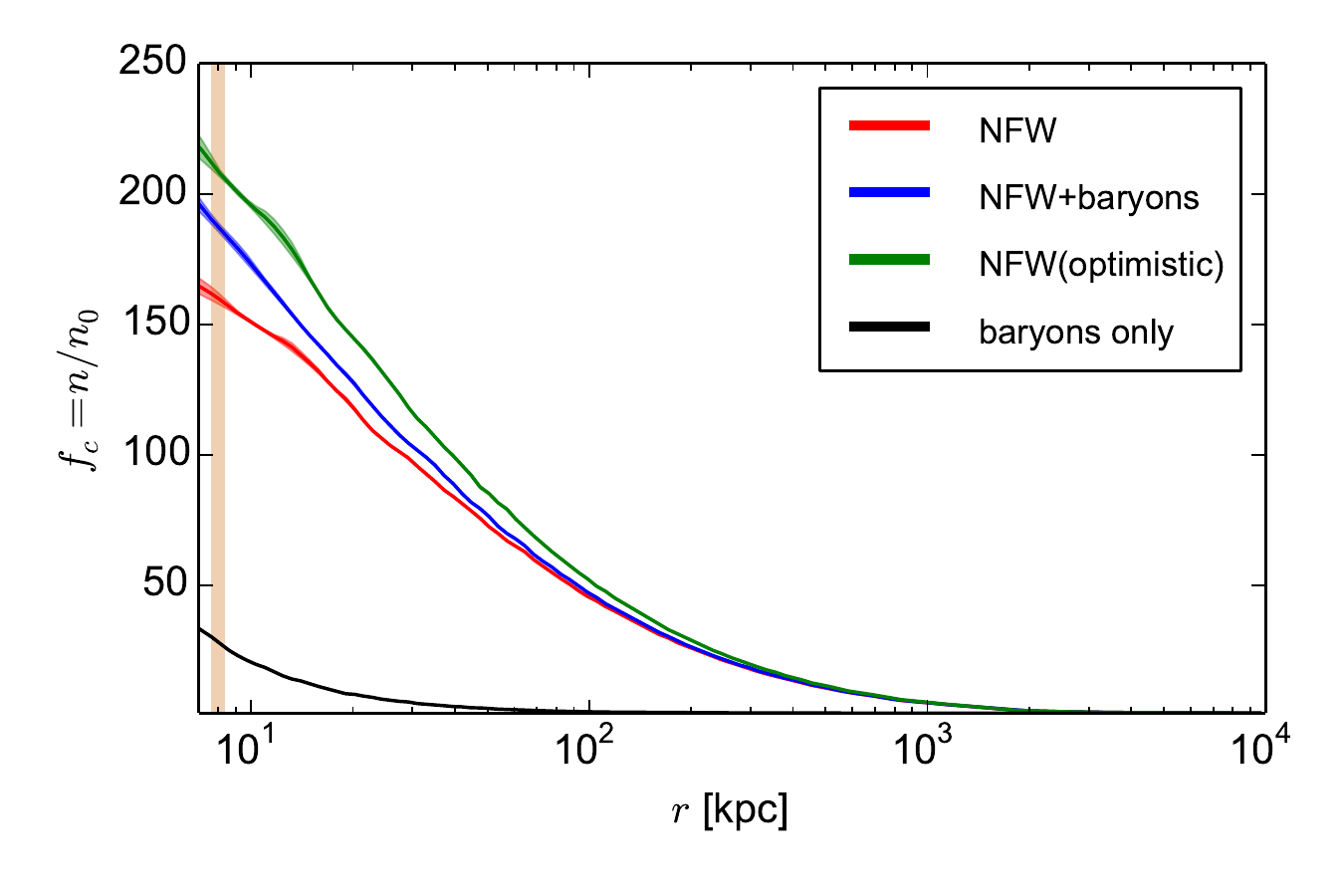}

\includegraphics[width=0.7\textwidth]{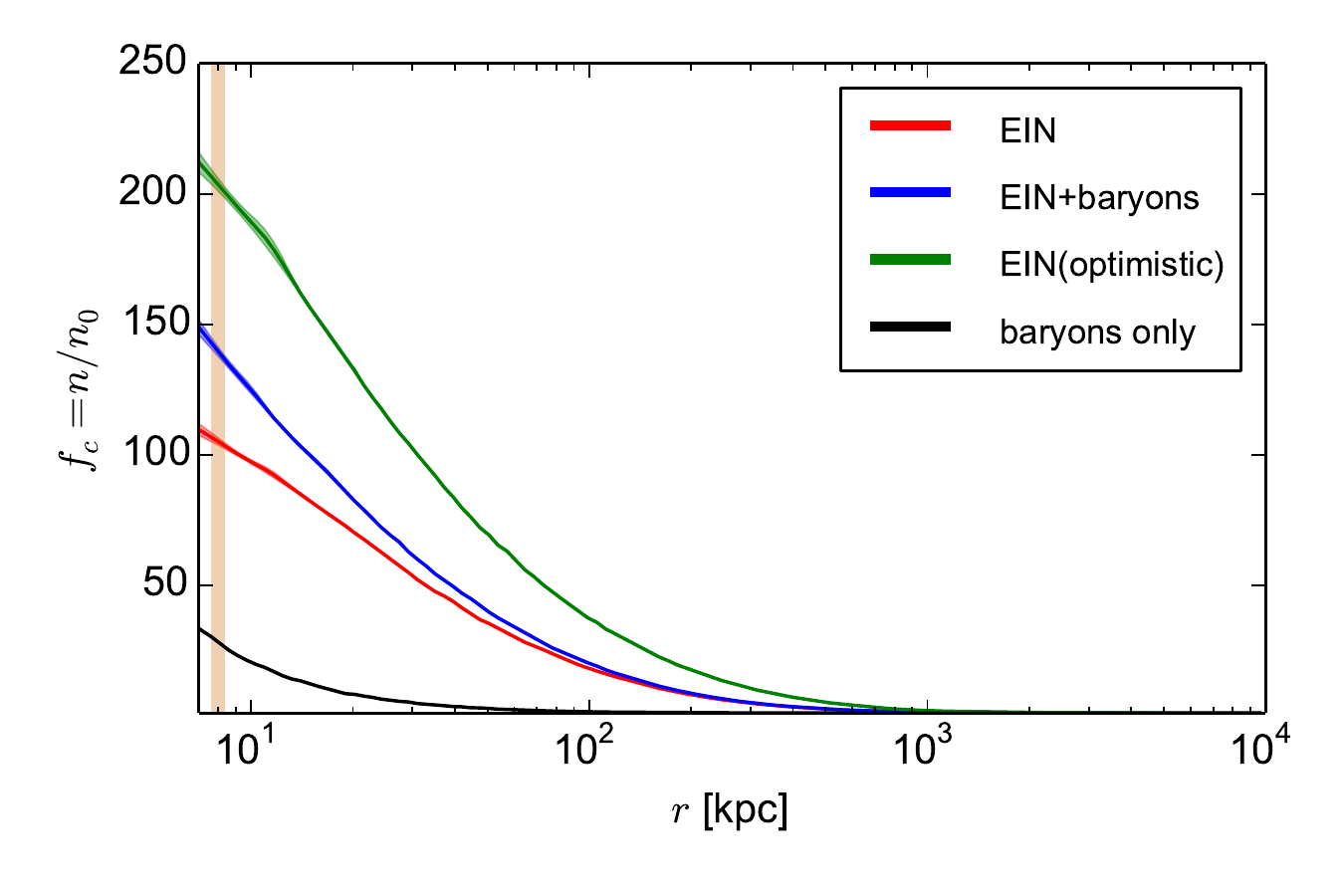}
\caption{
Same results as Figures \ref{fig:results60meV} and \ref{fig:results150meV}, but for a sterile neutrino with mass $m_4=1.3$ eV and contribution to the
radiation energy density $\DNeff = 1$.}
\label{fig:resultsSterile}
\end{figure}

In the following we will assume that $m_4\gg m_{1,2,3}$, so that we can approximate
$m_4\simeq\sqrt{\Delta m^2_{\rm SBL}}\simeq1.3\,\text{~eV}$.
This is the value for which we have calculated the clustering of sterile neutrinos in the local environment using \nob{} simulations.
The overdensity profiles for a neutrino with mass $m_4\simeq1.3$ eV are shown in figure\ \ref{fig:resultsSterile}.
As expected, the overdensity is much higher than in the previous cases because 
such neutrinos have a larger mass and smaller kinetic energies and 
they are more easily trapped by the galactic gravitational potential.

The total event rate for a Majorana (Dirac) fourth neutrino is given by \cite{Long:2014zva}
\begin{equation}\label{eq:nu4_eventrate}
 \Gamma_4^{M(D)}
 \simeq
 \DNeff\, |U_{e4}|^2\, f_c(m_4)\,\Gamma^{M(D)}_{\rm C\nu B}\,,
\end{equation}
from which we obtain values that vary between 3.4 and 33.9 (1.7 and 16.9),
depending on the assumptions on the matter profile of our galaxy
and on the thermalization of the fourth neutrino in the early Universe.
We list in table \ref{tab:ratesSterile} the expected event rates for two different values of $\DNeff$:
$\DNeff=1$, corresponding to a fully thermalized sterile neutrino (disfavoured by present cosmological constraints \cite{Ade:2015xua}),
and a conservative $\DNeff=0.2$, that is basically compatible with the 1$\sigma$ cosmological limits regardless of the assumed dataset.

\begin{table}[t]
\centering
\begin{tabular}{c|c||c|c|c}
matter halo &
overdensity $f_4$ &
$\DNeff$  &
$\Gamma_{\rm tot}^D$~(yr$^{-1}$)  &
$\Gamma_{\rm tot}^M$~(yr$^{-1}$) \\
\hline\hline
\multirowcell{ 2}{NFW(+bar)}      & \multirowcell{ 2}{159.9 (187.3)} & 0.2 &  2.6  (3.0) &  5.2  (6.1) \\
                                  &                                  & 1.0 & 13.0 (15.2) & 26.0 (30.4) \\
\hline
\multirowcell{ 2}{NFW optimistic} & \multirowcell{ 2}{208.6}         & 0.2 &  3.4        &  6.8        \\
                                  &                                  & 1.0 & 16.9        & 33.9        \\
\hline
\multirowcell{ 2}{EIN(+bar)}      & \multirowcell{ 2}{105.1 (139.5)} & 0.2 &  1.7  (2.3) &  3.4  (4.5) \\
                                  &                                  & 1.0 &  8.5 (11.3) & 17.1 (22.7) \\
\hline
\multirowcell{ 2}{EIN optimistic} & \multirowcell{ 2}{203.5}         & 0.2 &  3.3        &  6.6        \\
                                  &                                  & 1.0 & 16.5        & 33.0        \\
\hline
\end{tabular}
\caption{
Clustering factors and expected event rates in a PTOLEMY-like experiment
for different assumptions on the matter profile,
for a sterile neutrino corresponding to different values of $\DNeff$.
We consider a mass $m_4=1.3$~eV and a mixing matrix element $|U_{e4}|^2=0.02$~\cite{Gariazzo:2017fdh}.
}
\label{tab:ratesSterile}
\end{table}

\section{Conclusions}
\label{sec:conc}

The direct detection of cosmological relic neutrinos is one of the most challenging tasks of experimental astroparticle physics.
While other methods have been proposed in the literature (see e.g.~\cite{Domcke:2017aqj}), the most promising one taking into
account the possible values of neutrino masses is the capture of relic neutrinos on $\beta$-decaying nuclei. A first proposal
based on this technique, the PTOLEMY experiment \cite{Betts:2013uya}, is presently under development and the
corresponding expected number of events was studied in \cite{Long:2014zva}. The capture rate depends on the density of neutrinos
in our immediate vicinity, which is larger than the average cosmological density due to the attractive effect of our galaxy halo.

In this paper we have studied the gravitational clustering of massive neutrinos at galactic and subgalactic scales. In particular, we focused
on the matter distribution in the Milky Way, that we parameterised with different profiles for the dark matter and the baryons.
Using a method based on \nob{} simulations \cite{Ringwald:2004np}, we have computed the enhancement of the number density of relic neutrinos as
a function of the distance from the galactic centre for different values of the neutrino masses. For a 150~meV mass,
a value within the potential reach of PTOLEMY and not completely ruled out by present cosmological analyses including Planck data,
we find that the local density of cosmological neutrinos can be as large as two or three times its average value, depending on the
galactic matter profile. This result is in reasonable agreement with previous analyses \cite{Ringwald:2004np,VillaescusaNavarro:2012ag}.
We have also considered, for the first time, smaller neutrino masses. A nearly minimal value for the heaviest neutrino is 60~meV, for which
we find that the local population of relic neutrinos is also enhanced, but only up to $10$-$20\%$ with respect to its average number density.

These very small enhancement factors can be compared with the significant overdensities that could be reached if neutrino masses were larger than 1 eV, a case
that is not possible for standard, active neutrinos. As a non-standard example, we have considered the case of a fourth massive state, mostly sterile, introduced
in order to provide an explanation to the short-baseline neutrino oscillation anomalies. This possibility, however, is disfavoured by cosmological data and, in any case, would require that these heavier neutrinos were produced in the early Universe but not completely thermalized. Nevertheless, we find that gravitational clustering would
lead to enhancement factors of order $140-210$ for a $1.3$ eV neutrino mass.

After computing the local neutrino overdensities for each case, we have calculated the consequences for a PTOLEMY-like experiment.
The corresponding event rates depend on whether neutrinos are Dirac or Majorana particles, as well as on the neutrino mass ordering (which fixes the mixing
of the mass eigenstates with the electron flavour). In the case of sterile neutrinos, the number of events also depends on their degree of thermalization at production and
their mixing with electron neutrinos. In any case, a positive detection of the relic neutrinos at PTOLEMY will be very difficult for the smallest neutrino masses,
unless its energy resolution is significantly improved.

The main uncertainties in the calculation of our results on massive neutrino clustering in the Milky Way are related to the parameterisation of the galactic matter distribution, both for baryons and DM. We found that the differences obtained when modifying the matter content are much larger than the numerical uncertainties in the \nob{} calculations.
In the near future, an eventual measurement of the absolute scale of neutrino masses, their Dirac or Majorana nature and a better knowledge of the distribution
of matter in our galaxy will lead to an improved calculation of the local overdensity of relic neutrinos and a better prediction of the event rate at a PTOLEMY-like experiment.

\acknowledgments
We thank E.~Castorina, R.~Lineros, M.~Viel and F.~Villaescusa-Navarro for fruitful discussions on N-body simulations and the matter content of our galaxy.
P.F.\ de Salas thanks the Aachen  Institute for Theoretical Particle Physics and Cosmology (TTK) for hospitality and support when this work began.
Work supported by the Spanish grants FPA2014-58183-P, FPA2015-68783-REDT, Multidark CSD2009-00064 and SEV-2014-0398
(MINECO), FPU13/03729 (MECD) and PROMETEOII/2014/084 (Generalitat Valenciana).


\end{document}